# Subject-Level Heterogeneity in EEG Motor Imagery Decoding: A Large-Scale Benchmark and Portfolio-Based Reduction of the Search Space

Xavier Vasques [1, 2], Paul Barbaste [6, 7], Olivier Oullier [3, 4, 5*]

[1] *IBM Technology,* Bois-Colombes, France
[2] *IBM France Lab,* Orsay, France
[3] *Inclusive Brains,* Marseille, France
[4] *Human-Computer Interaction Department, Mohamed bin Zayed University of Artificial Intelligence,* Abu Dhabi, UAE
[5] *Institute for Artificial Intelligence,* Biotech Dental Group, Salon-de-Provence, France
[6] *Wavestone,* Paris, France
[7] *Human Technology Foundation,* Paris, France

*Corresponding author:
Olivier Oullier olivier.oullier@mbzuai.ac.ae

**ABSTRACT**

Robust EEG motor imagery decoding remains limited by strong inter-individual variability, making it difficult to identify pipelines that generalize reliably across users. Here, we present a large-scale, standardized within-session benchmark of EEG motor imagery decoding pipelines across three public datasets: Cho2017 (52 subjects), PhysionetMI (109 subjects), and Zhou2016 (4 subjects). Using a common MOABB LeftRightImagery setting, two frequency bands (8–15 Hz and 8–30 Hz), and a broad combination of feature extraction, preprocessing, and classification steps, we analyzed 216,714 raw evaluation pipelines, corresponding after aggregation to 61464 subject-level observations in Cho2017, 132762 in PhysionetMI, and 22488 in Zhou2016.

At the benchmark level, covariance tangent-space projection (cov-tgsp) and Common Spatial Patterns (CSP) consistently defined the strongest methodological families, although their relative ordering was dataset-dependent. On Cho2017, the best family-level mean subject-level accuracy was obtained with cov-tgsp in 8–30 Hz (0.712 ± 0.140), whereas Zhou2016 favored CSP (0.832 ± 0.121 in 8–15 Hz). PhysionetMI showed a flatter upper ranking, with cov-tgsp and CSP nearly tied (0.605 ± 0.174 vs 0.604 ± 0.178 in 8–30 Hz). However, these aggregate rankings concealed substantial subject-level heterogeneity: 42 distinct winning pipelines were observed across 52 Cho2017 subjects and 93 across 109 PhysionetMI subjects.

To exploit this heterogeneity, we used the exhaustive benchmark as an empirical performance landscape for constructing compact portfolios of candidate pipelines, where $K$ denotes portfolio size. Several portfolio-construction procedures were compared, including the ranking-based Top-K Mean heuristic and search-based strategies such as Greedy, Greedy-Diverse, Beam Search, Regret-Greedy, Submodular Coverage, and Family-Constrained selection. Across strategies, results were broadly consistent, and Top-K Mean provided the strongest overall trade-off. Under this approach, a single best global pipeline already retained 94.2% of the oracle in Cho2017 and 81.8% in PhysionetMI; increasing portfolio size to $K = 12$ raised oracle retention to 96.5% and 90.0%, respectively. These findings show that the decoding landscape is subject-dependent, and that this heterogeneity can be exploited through compact portfolios that make future personalization more feasible.

## INTRODUCTION

Brain-computer interfaces (BCIs) have attracted growing attention as both invasive [1] and non-invasive [2] technologies for enabling communication and control through brain activity and external devices [3]. In clinical contexts, BCIs may help individuals with severe motor impairments, including locked-in syndrome, amyotrophic lateral sclerosis, spinal cord injury, or disabling speech and movement disorders following stroke, to communicate and sometimes regain forms of communication or device control [4–8] [9]. More broadly, BCIs rely on decoding neural activity into actionable commands, allowing users to interact with computers, assistive devices, prostheses, or communication systems [6,10–12].

Among non-invasive neurotechnologies, electroencephalography (EEG) remains the most widely used platform for BCI research and development [2]. Its main advantages are safety, affordability, portability, and high temporal resolution. However, decoding mental states such as motor imagery from EEG remains difficult because brain signals are intrinsically variable [13] and because non-invasive recordings are noisy and vulnerable to multiple sources of contamination [6,14,15]. Motor imagery itself, although known to share neural substrates with overt movement [16], is therefore not straightforward to decode robustly across individuals and recording conditions.

A major obstacle is the strong inter- and intra-individual variability of EEG signals [17]. Signal properties can vary substantially across users, and also across sessions recorded from the same user, because of differences in attention, fatigue, electrode placement, impedance, physiology, and environmental conditions [3]. This non-stationarity complicates the development of generalized decoding pipelines and remains a central barrier to the reliable deployment of BCIs in real-world medical and everyday settings [6]. The phenomenon often referred to as "BCI illiteracy" [11,18,19], affecting a substantial proportion of users, further highlights the need for decoding strategies that are both more robust and more adaptive [20,21].

Machine learning (ML) and deep learning (DL) methods have been widely explored to address these challenges [22–24]. Classical methods such as support vector machines (SVMs) [25] and linear discriminant analysis (LDA) [26] remain central in EEG classification, while convolutional neural networks (CNNs) and related architectures have expanded the DL literature [27,28]. At the feature level, spatially informed approaches have played a major role in motor imagery decoding, particularly Common Spatial Patterns (CSP) [29] and Riemannian geometry-based methods operating on trial covariance matrices with tangent-space projections [30,31] [32]. These methods often achieve strong performance, but they may generalize poorly when inter-subject variability becomes prominent or when recording conditions shift [33].

For this reason, evaluating a broader range of EEG representations has become increasingly important. EEG feature extraction spans a wide spectrum of approaches, from time- and frequency-domain summaries to more complex multidimensional representations [34–36]. In addition, generalization across users is likely to depend on the availability of robust and transferable features, especially in the presence of dataset shift and non-stationarity [37]. Beyond spatial methods, nonlinear descriptors may capture aspects of EEG complexity that are more closely related to fluctuating subject states, such as fatigue or variable engagement. Examples include Hjorth parameters, Higuchi's fractal dimension (HFD), Hurst exponent, Fisher-information-based measures, and singular value decomposition entropy (SVD) [14,38–40]. Functional-connectivity measures provide another complementary perspective by characterizing task-related interactions between EEG channels and have increasingly been considered in motor imagery decoding pipelines [41–43].

These developments motivate a broader question than simply identifying the single best decoding method on average: how does decoding performance vary across datasets, methodological families, and individual subjects, and can this variability be exploited constructively? Addressing this question requires evaluation procedures that are both rigorous and reproducible, including careful control of leakage, appropriate performance metrics, and standardized comparison frameworks [44–47]. It also requires going beyond purely aggregated group-level summaries [48], because average rankings may conceal marked subject-specific differences in the identity of the best-performing pipeline.

In the present study, we therefore performed a large-scale, standardized benchmark of EEG motor imagery decoding pipelines across three publicly available datasets widely used in BCI research: the PhysioNet EEG Motor Movement/Imagery dataset [49], the Zhou2016 dataset [40], and the Cho2017 dataset [50]. The benchmark included 216,714 unique combinations of feature extraction, preprocessing, and classification procedures, spanning spatial filters such as CSP [51], covariance-based and tangent-space approaches [46,52,53], functional-connectivity representations [54–56], and several nonlinear descriptors including Hjorth parameters [57], SVD entropy [58], Hurst exponent [38], Fisher information [59], and HFD [60]. To ensure comparability and reproducibility, all evaluations were conducted within the Mother of All BCI Benchmarks (MOABB) framework [61,62] under a harmonized within-session LeftRightImagery setting.

Our objective, however, is not limited to ranking pipelines by average performance. The first aim of this work is to characterize the performance landscape of motor imagery decoding at multiple complementary levels: overall benchmark structure, methodological-family effects, dataset-specific differences, and subject-level heterogeneity. The second aim is to determine whether the exhaustive benchmark itself can be used as an empirical basis for practical pipeline reduction. In other words, rather than assuming that a single universally optimal configuration exists, we ask whether compact portfolios of pipelines, selected from the exhaustive benchmark, can retain most of the subject-specific oracle while substantially reducing the number of candidate pipelines that would need to be tested in practice.

This benchmark-driven perspective extends the role of exhaustive benchmarking. Rather than treating it solely as a descriptive exercise, we use it to identify globally competitive methodological families, quantify how strongly the optimal configuration depends on the subject, and test whether this heterogeneity can be exploited through compact portfolio selection. In this way, the present study provides both a systematic characterization of large-scale EEG motor imagery decoding performance and a practical framework for reducing the search space toward future subject-aware pipeline selection strategies.

# MATERIALS & METHODS

## Datasets

To benchmark EEG motor imagery decoding pipelines under a standardized evaluation framework, we used three publicly available datasets through the Mother Of All BCI Benchmarks (MOABB) framework introduced by Jayaram et al. [62,63]: Physionet Motor Imagery, Zhou2016, and Cho2017. The PhysionetMI dataset includes recordings from 109 participants acquired with a 64-channel EEG system at 160 Hz [64]. The original protocol comprised multiple runs including baseline, executed movement, and imagined movement tasks. For the present study, we retained the motor imagery conditions corresponding to left- versus right-hand imagery and integrated this dataset into the MOABB framework using the BCI2000 documentation and event annotations. The Zhou2016 [65] dataset includes 4 participants recorded with 14 EEG channels at 250 Hz across three sessions. The original protocol involved three motor imagery classes (left hand, right hand, and feet), with repeated trials collected over sessions separated by days to months. For consistency with the present benchmark, only the left- versus right-hand imagery (LeftRightImagery) conditions were considered. Because this dataset includes only four participants, results derived from it were interpreted cautiously and not given the same weight as those from larger datasets. The Cho2017 [14] dataset includes 52 participants recorded with 64 EEG channels at 512 Hz using a BioSemi ActiveTwo system. The experiment was designed to study left- and right-hand motor imagery, with several runs per participant and brief rest periods between runs. To standardize signal extraction across datasets, we used two commonly adopted motor imagery frequency ranges (8-15 Hz and 8-30 Hz) and a post-cue time window of 0.6-2.0 s. Dataset-specific sampling frequencies were preserved as provided in the original recordings (160 Hz for PhysionetMI, 250 Hz for Zhou2016, and 512 Hz for Cho2017). This common MOABB-based configuration enabled direct comparison of feature extraction and classification pipelines under a consistent within-session evaluation framework across datasets with different cohort sizes and acquisition characteristics.

## EEG Signal Processing

### Filtering

EEG signals were band-pass filtered using two commonly adopted motor imagery frequency ranges: 8-15 Hz and 8-30 Hz. The 8-15 Hz range was selected to focus primarily on the mu rhythm, whereas the 8-30 Hz range was used to capture a broader sensorimotor spectrum including both mu and beta activity.

For epoch extraction, analyses were restricted to a post-cue time window of 0.6-2.0s. This time interval was chosen to focus on the motor imagery period while reducing the influence of the earliest post-stimulus evoked responses. The same temporal window and frequency settings were applied consistently across datasets within the MOABB benchmarking configuration.

### EEG Feature Extraction

We evaluated four broad families of EEG feature extraction methods: (i) spatial filtering using Common Spatial Patterns (CSP), (ii) covariance-based tangent-space representations derived from Riemannian geometry, (iii) coherence-based tangent-space representations intended to capture functional connectivity patterns, and (iv) non-linear time-series descriptors. These feature families were selected because they reflect commonly used yet methodologically distinct approaches for motor imagery decoding. Unless otherwise specified, feature extraction was applied to the band-pass filtered post-cue epochs described above.

CSP, described by Koles et al. [66], is a standard spatial filtering approach for two-class EEG decoding that identifies linear projections maximizing variance differences between classes. In this study, CSP was applied directly to the left- versus right-hand motor imagery epochs in order to derive low-dimensional spatial features for classification. We did not use filter-bank CSP; instead, CSP was computed separately within each of the two predefined frequency ranges (8-15 Hz and 8-30 Hz). The resulting CSP-transformed signals were then used as inputs to the downstream classifiers.

We also evaluated covariance-based Riemannian representations [46,67]. For each epoch, channel covariance matrices were computed and projected into tangent space at the Riemannian geometric mean reference point. This yielded Euclidean feature vectors derived from symmetric positive definite covariance matrices, enabling the use of conventional classifiers while preserving information about the covariance structure of multichannel EEG signals. As the dimensionality of this representation grows with the number of channels, the resulting feature space was substantially higher-dimensional than that of CSP.

To capture functional interactions between channels, we also computed coherence-based connectivity matrices from each epoch and projected them into tangent space using the same general Riemannian framework. These features were intended to summarize pairwise interaction patterns between EEG channels in a form compatible with Euclidean classifiers. This approach provided an additional representation complementary to covariance-based spatial features.

In addition to spatial and covariance-based approaches, we extracted several non-linear descriptors intended to characterize signal complexity and temporal irregularity. These included the Higuchi fractal dimension, Hjorth mobility and complexity [15], singular value decomposition (SVD) entropy, and the Hurst exponent. For Higuchi fractal dimension, the maximum scale parameter was set to $k = 10$. Hjorth parameters were computed directly from each time series, and SVD entropy and Hurst exponent were calculated using their standard formulations. These features were extracted independently for each channel and then concatenated prior to classification.

Because the dimensionality of the extracted feature spaces differed substantially across methods, the compared pipelines should be interpreted as end-to-end decoding configurations rather than as strictly dimension-matched feature representations. This point is important when comparing low-dimensional methods such as CSP with higher-dimensional tangent-space approaches.

## Data preprocessing

As part of the benchmark, we evaluated several feature rescaling and transformation strategies before classification. These preprocessing steps were applied to the extracted feature vectors and treated as components of the decoding pipelines rather than as a separate optimization stage.

Their inclusion was motivated by the fact that EEG-derived feature spaces can differ substantially in scale, dispersion, and distribution across methods.

The evaluated preprocessing methods included StandardScaler, MinMaxScaler, MaxAbsScaler, RobustScaler, L2 normalization, sigmoid-based transformation, power transformation (Yeo-Johnson), and quantile-based transformations. These transformations were used to test how common normalization and distribution-shaping procedures affect downstream classification performance across heterogeneous feature representations.

Unless otherwise specified, preprocessing was applied after feature extraction and before model fitting using standard implementations from the corresponding Python machine learning libraries. Because preprocessing choice was part of the benchmarked pipeline definition, each scaler/transformation contributed to the total number of evaluated configurations.

## Machine Learning Classifiers

We evaluated several classical machine learning classifiers on the extracted EEG feature representations. These classifiers were combined with the different preprocessing and feature extraction steps described above to form complete decoding pipelines.

The benchmark included logistic regression with elastic-net regularization, random forest, linear support vector machine (linear SVM), radial basis function support vector machine (RBF-SVM), and linear discriminant analysis (LDA). These models were selected because they span a range of commonly used linear, kernel-based, ensemble, and probabilistic approaches for EEG decoding.

In addition, we evaluated a predefined set of 17 multilayer perceptron (MLP) architectures, indexed as MLP_1 to MLP_17, in order to assess how shallow feedforward neural classifiers behaved across heterogeneous feature spaces. These architectures differed in the number of hidden layers, number of hidden units, activation functions, optimization solvers, and regularization settings. The explored configurations included both simple single-hidden-layer models and deeper architectures, with training based on either stochastic gradient descent or Adam optimization. L2 regularization was controlled through the alpha parameter.

Unless otherwise specified, classifier hyperparameters were fixed within each configuration and applied consistently across datasets and feature families. In this study, classifiers were therefore treated as predefined components of the benchmarked pipelines rather than as the result of a separate hyperparameter optimization procedure.

For reproducibility and traceability, each classifier configuration, including each MLP variant, was assigned a unique identifier and used consistently throughout the benchmark.

## Assessment

Pipeline performance was evaluated using the within-session evaluation protocol implemented in the MOABB framework, following the standardized benchmarking approach described by Jayaram and Barachant [62]. Under this protocol, each subject's data are partitioned into predefined training and test subsets within the same recording session, and all evaluated pipelines are trained and tested on identical splits for a given dataset and paradigm.

For each dataset, frequency band, and feature configuration, each pipeline was fitted on the training subset and evaluated on the corresponding held-out test subset, yielding one performance profile per subject. These subject-level results were then aggregated across participants to summarize the central tendency and dispersion of each pipeline. All analyses were conducted independently for PhysionetMI, Zhou2016, and Cho2017, and for both frequency ranges considered in this study (8-15 Hz and 8-30 Hz).

The primary benchmark metric was accuracy, computed at the subject level and then aggregated across subjects for each pipeline. To provide a broader and more robust assessment of classification performance, we also evaluated balanced accuracy, ROC-AUC, F1-score, precision, and recall. Performance dispersion across subjects was characterized using summary statistics including mean, standard deviation, median, and interquartile range.

To complement descriptive rankings, inferential statistical analyses were planned on subject-level scores for selected benchmark comparisons. Global differences between pipelines were to be assessed using non-parametric repeated-measures tests, followed where appropriate by post-hoc pairwise comparisons with correction for multiple testing. Effect sizes were also to be reported whenever feasible.

In addition to pipeline-level rankings, results were to be summarized at the level of broader methodological families, including feature extraction family, classifier family, frequency range, and dataset. This family-level analysis was intended to provide a more interpretable view of performance trends beyond individual pipeline variants. To further characterize inter-subject heterogeneity, we also planned to quantify how often each feature family achieved the best subject-level performance within each dataset.

Because the dimensionality of the extracted feature representations differed substantially across methods, an additional analysis was planned to document feature dimensionality and examine its relationship to classification performance. This analysis was intended to support interpretation of comparisons between low-dimensional approaches such as CSP and higher-dimensional tangent-space representations.

Finally, the full benchmark design was to be documented through an explicit decomposition of the evaluated configurations, including datasets, frequency bands, feature extraction methods, preprocessing transformations and classifier families.

Importantly, this within-session protocol evaluates generalization to unseen trials from the same recording session only. It does not assess cross-session robustness or participant-independent transfer. Accordingly, the results reported here should be interpreted as a standardized within-session comparison of decoding pipelines rather than as evidence of subject-independent generalization.

## Portfolio construction from the exhaustive benchmark

To complement the exhaustive benchmark, we performed a benchmark-driven portfolio analysis aimed at identifying small global sets of candidate pipelines that could retain most of the subject-specific oracle while substantially reducing the number of pipelines that would need to be considered in practice. Importantly, this analysis did not attempt to assign a different pipeline to each subject from subject-level EEG features. Instead, it treated the exhaustive benchmark itself as an empirical performance landscape from which compact portfolios of pipelines could be constructed and evaluated. For each dataset and frequency band, the benchmark results were organized into a subject-by-pipeline performance matrix, in which each row corresponded to a subject, each column corresponded to a candidate pipeline, and each cell contained the subject-level performance of that pipeline under the selected metric. In the present study, portfolio construction was based on balanced accuracy. This matrix was the only input used for portfolio selection. We evaluated portfolio sizes (K = 2, 3, 4, 6, 8, 10,) and (12). For interpretability, (K=1) was additionally defined in the Results section as the single best global pipeline, i.e. the pipeline with the highest mean subject-level performance within a given dataset-band configuration. For (K>1), the objective was to select a subset of (K) pipelines that would collectively preserve as much of the observed subject-level heterogeneity as possible.

Portfolio construction was performed using repeated subject-level train-test splits within each dataset-band combination. Specifically, subjects were split using ShuffleSplit with 10 repetitions, a test fraction of 0.2, and a fixed random seed of 42. This corresponds to repeated 80/20 train-test splits at the subject level, so that approximately 80% of subjects were used to construct the portfolio and 20% were reserved for evaluation. The portfolio was always learned on the training subjects only and then evaluated on the held-out test subjects. This design was used to estimate how well compact global portfolios generalized to unseen subjects within the same dataset and frequency-band configuration.

Several heuristic portfolio-construction strategies were compared: Top-K Mean (*topk_mean*), Greedy (*greedy*), Greedy-Diverse (*greedy_diverse*), Beam Search (*beam*), Regret-Greedy (*regret_greedy*), Submodular Coverage(*submodular_coverage*), and Family-Constrained Selection (*family_constrained*). All strategies operated on the same training subject-by-pipeline matrix, but they differed in how they selected candidate subsets.

Top-K Mean selected the (K) pipelines with the highest mean training performance. It therefore favored pipelines that were strongest on average, without explicitly modeling complementarity. Greedy selection was iterative: starting from an empty set, it added at each step the pipeline that most improved the current portfolio. Greedy-Diverse followed the same iterative procedure but penalized pipelines that were too redundant with those already selected. Beam Search extended the greedy idea by keeping several promising partial portfolios at each step rather than only one. Regret-Greedy prioritized reduction of the gap between the current portfolio and the full benchmark oracle. Submodular Coverage emphasized marginal coverage gain while also encouraging diversity. Family-Constrained Selection imposed an initial diversity constraint at the methodological-family level before completing the portfolio greedily.

For the strategies based on iterative or objective-driven search, candidate portfolios were scored using a composite objective designed to favor both performance and complementarity. This objective combined: (i) the mean oracle performance recoverable within the selected set, (ii) the gain over the best single fixed pipeline contained in that set, (iii) a penalty for redundancy between selected pipelines based on positive pairwise correlations of their subject-level performance profiles, and (iv) a penalty for residual regret relative to the full benchmark oracle. In this way, the portfolio-construction procedure favored subsets composed of pipelines that were both strong on average and useful on different subsets of subjects, rather than collections of highly similar pipelines.

Portfolio performance on the held-out test subjects was summarized using four related quantities. First, the global oracle was defined as the mean, across test subjects, of the best available pipeline among all benchmarked pipelines. This quantity represents the maximal subject-specific benchmark performance achievable if the best pipeline could always be chosen retrospectively for each subject. Second, the oracle-in-set was defined as the mean, across test subjects, of the best score achieved by any pipeline contained in the selected portfolio. This quantity represents the best performance that the portfolio could offer if, for each test subject, the most suitable pipeline within that portfolio could be chosen.

Third, the fixed best global pipeline was defined as the single pipeline with the highest mean test performance across subjects when considering all benchmarked pipelines. This corresponds to the strongest one-size-fits-all solution available in the full benchmark. Fourth, the fixed-best-in-set was defined as the single pipeline within the selected portfolio with the highest mean test performance across the held-out subjects. This corresponds to the strongest one-size-fits-all solution available within that portfolio.

From these quantities, we derived two summary metrics. The first was the oracle-retention ratio, defined as

$$oracle - retention\, ratio = \frac{oracle - in - set}{global\ oracle}$$

This ratio quantifies how much of the full subject-specific oracle is preserved by a compact portfolio. A value close to 1 indicates that the selected portfolio retains most of the maximal subject-level performance available in the full benchmark. The second was the absolute gain relative to (K=1), computed as the difference between the oracle-in-set of a portfolio of size (K) and the performance of the best single global pipeline ((K=1)). This quantity measures the practical improvement obtained by moving from a single globally fixed pipeline to a compact portfolio.

Finally, in addition to performance evaluation, we also examined the methodological-family composition of the selected portfolios. This made it possible to determine whether useful diversity arose mainly across feature families or rather within the dominant family already identified by the benchmark. Cross-validated portfolio summaries were then aggregated by dataset, frequency band, portfolio size, and search strategy for downstream analysis and visualization.

## Code availability

All analyses were implemented in Python 3 using standard open-source scientific computing libraries. Classical machine learning pipelines were developed primarily with scikit-learn [68], Riemannian feature extraction relied on PyRiemann, non-linear feature extraction used PyEEG, and EEG preprocessing and handling were performed with MNE-Python. Dataset loading and benchmark evaluation were conducted through the MOABB framework.

The code used to run the benchmark and reproduce the analyses is available on GitHub at: https://github.com/xaviervasques/EEG_benchmark/tree/main. Additional implementation details, including package dependencies and configuration settings, are provided in the repository.

# RESULTS

In this study, results are reported for a standardized within-session benchmark conducted across three public motor imagery EEG datasets: PhysionetMI [69], Cho2017 [14], and Zhou2016 [65]. All evaluated pipelines were assessed within the MOABB framework using the same LeftRightImagery paradigm, allowing direct comparison across datasets despite differences in cohort size, sampling frequency, and electrode configuration [62,63].

The benchmark covered two commonly used motor imagery frequency ranges, 8-15 Hz and 8-30 Hz, and included several methodological families: the covariance projection method in tangent space (cov-tgsp), a state-of-the-art approach[46], with other techniques such as common spatial patterns on covariances (csp)[47], functional connectivity projected onto tangent space (con_instantaneous + tgsp)[70], and non-linear features including Hjorth parameters (hjorth)[15], Higuchi fractal dimension (hfd)[39], and singular value decomposition entropy (svd_entropy)[71]. For cov-tgsp and con_instantaneous+tgsp, 2080 features were extracted, 64 for svd_entropy and hfd, 192 for Hjorth and two components for CSP. These feature extraction methods were combined with multiple preprocessing strategies and classifier families, generating a large collection of decoding pipelines evaluated under the same experimental framework.

Because the present study relies on within-session evaluation, the reported results should be interpreted as a comparison of subject-level decoding performance on unseen trials from the same recording session, rather than as evidence of cross-subject or cross-session generalization. Accordingly, the main objective of the Results section is not to claim participant-independent transfer, but to characterize how pipeline performance varies across datasets, frequency bands, methodological families, and individual subjects under a controlled and reproducible benchmarking setting.

To support this analysis, results are presented at four complementary levels: overall benchmark trends, methodological family comparisons, dataset-specific patterns, and subject-level heterogeneity. Although exhaustive benchmarking offers a detailed and reproducible characterization of the performance landscape, it remains computationally demanding and of limited practical use if thousands of pipelines must be tested for each new user. We therefore used the exhaustive benchmark as an empirical performance landscape from which compact portfolios of pipelines could be selected, in order to assess whether most of the subject-specific oracle could be retained while substantially reducing the practical search space.

## Global performance landscape across datasets, methodological families, and frequency bands

The consolidated benchmark comprised 216714 raw evaluation rows across the three datasets, including 61464 rows for Cho2017, 132762 for PhysionetMI, and 22488 for Zhou2016. Because the raw outputs contained repeated entries at the row level, all benchmark-level summaries were computed after structured aggregation. Repeated observations were first collapsed at the subject-session-pipeline level and, when applicable, then averaged across sessions to derive a single subject-level score for each pipeline. This yielded 44928 subject-level pipeline observations for Cho2017, 109000 for PhysionetMI, and 4192 for Zhou2016. A full summary of benchmark volume and subject-level performance metrics across datasets, methodological families, and frequency bands is provided in **Table 1**, while the overall distribution of subject-level scores is illustrated in **Figure 1**. The whole benchmark results are provided in supplementary material (**supplementary_material_benchmark**).

At the broadest level, the benchmark revealed a non-uniform performance landscape structured by methodological family, dataset, and frequency band. Two families consistently occupied the leading positions: cov-tgsp and CSP. In contrast, coherence-based tangent-space representations and the nonlinear descriptors remained secondary across most benchmark settings. This global hierarchy was visible both in the aggregated subject-level statistics reported in **Table 1** and in the score distributions shown in **Figure 1**, indicating that the benchmark was not characterized by a flat performance profile across decoding strategies.

A first notable pattern concerned the comparison between the two leading spatially informed families (**Figure 2**). On Cho2017, the strongest family-level result was obtained with cov-tgsp in the 8-30 Hz band, which reached a mean subject-level accuracy of 0.712 ± 0.140, clearly ahead of CSP in 8-15 Hz (0.667 ± 0.156) and 8-30 Hz (0.661 ± 0.158). The same dataset also showed the clearest separation between the leading spatial families and the remaining approaches. Coherence-based tangent-space representations reached 0.630 ± 0.115 in 8-15 Hz and 0.604 ± 0.118 in 8-30 Hz, while the nonlinear families remained below 0.60, with performances ranging from 0.533 ± 0.065 for HFD in 8-15 Hz to 0.586 ± 0.092 for HFD in 8-30 Hz. Thus, on Cho2017, the performance hierarchy was well defined and favored covariance-based tangent-space features.

The pattern was more compressed on PhysionetMI, where the difference between the two leading families became much smaller. The best mean subject-level accuracy was again obtained with cov-tgsp in 8-30 Hz (0.605 ± 0.174), but this result was only marginally higher than CSP in 8-30 Hz (0.604 ± 0.178) and CSP in 8-15 Hz (0.596 ± 0.172). This near-equivalence between cov-tgsp and CSP distinguishes PhysionetMI from Cho2017 and suggests a flatter upper part of the ranking in the most heterogeneous dataset of the benchmark. Below these two leading families, performances decreased but remained relatively tightly grouped: nonlinear descriptors ranged between 0.527 ± 0.113 and 0.534 ± 0.119, while the weakest result on this dataset was obtained with con_instantaneous+tgsp in 8-30 Hz (0.495 ± 0.118). Compared with Cho2017, PhysionetMI therefore exhibited both lower absolute performance and reduced separation between the top-ranking families.

A third configuration emerged on Zhou2016, where CSP became the dominant family. Mean subject-level accuracy reached 0.832 ± 0.121 in 8-15 Hz and 0.818 ± 0.138 in 8-30 Hz, clearly above cov-tgsp, which ranked second with 0.752 ± 0.112 in 8-15 Hz and 0.756 ± 0.115 in 8-30 Hz. Although Zhou2016 contains only four subjects and therefore must be interpreted cautiously, the ranking reversal relative to Cho2017 is noteworthy. It indicates that the benchmark-level hierarchy is not fixed and that the dominant family depends on dataset characteristics. The remaining families again occupied lower positions, with nonlinear descriptors ranging from 0.593 ± 0.085 for Hjorth in 8-30 Hz to 0.646 ± 0.087for HFD in 8-30 Hz.

Beyond the dataset effect, the benchmark also revealed family-specific sensitivity to frequency band (**Figure 2**). Among the leading families, cov-tgsp showed a clear preference for the broader 8-30 Hz setting on Cho2017 and PhysionetMI, where its best results were always obtained in that band. By contrast, CSP displayed weaker band sensitivity overall, with relatively similar values between 8-15 Hz and 8-30 Hz on Cho2017 and PhysionetMI, but a clear advantage for 8-15 Hz on Zhou2016. The coherence-based tangent-space family did not benefit consistently from the broader band and even showed its weakest performance on PhysionetMI in 8-30 Hz. Among the nonlinear descriptors, the broad-band setting tended to improve HFD, particularly on Cho2017 and Zhou2016, whereas the gains for Hjorth and SVD entropy were smaller and less systematic. Taken together, these results suggest that the effect of frequency band was not uniform across methodological families and should be interpreted as an interaction with both feature representation and dataset.

Pipeline-level rankings were consistent with these family-level patterns. On Cho2017, the best-performing pipeline combined cov-tgsp in the 8-30 Hz band with MinMax scaling and a linear SVM, reaching a mean subject-level accuracy of 0.738 ± 0.132. It was closely followed by the same feature representation and scaling strategy combined with elastic-net logistic regression, which achieved 0.736 ± 0.143. On PhysionetMI, the highest mean score was obtained with cov-tgsp (8-30 Hz) combined with RobustScaler and a linear SVM, yielding 0.650 ± 0.183, followed by the same feature representation combined with StandardScaler and a linear SVM (0.639 ± 0.185). On Zhou2016, the best-performing pipeline used CSP in the 8-15 Hz band, followed by MinMax scaling and a MLP, reaching 0.881 ± 0.074. Thus, the pipeline-level results did not contradict the family-level patterns, but rather refined them by showing that the strongest configurations were consistently nested within the two leading methodological families.

Overall, the benchmark identified cov-tgsp and CSP as the only families that remained competitive across all three datasets, although their relative ordering depended on both dataset and frequency band. Cho2017 favored covariance tangent-space representations, Zhou2016 favored CSP, and PhysionetMI placed the two families in near-equilibrium at the top of the ranking. The remaining approaches, including coherence-based tangent-space representations and nonlinear descriptors, contributed useful contrast within the benchmark but did not challenge the two leading families at the global level.

**Table 1.** Global benchmark structure and subject-level performance summaries across datasets, methodological families, and frequency bands. The table summarizes the overall structure of the benchmark and the aggregated subject-level performance obtained for each dataset, methodological family, and frequency band. Reported values correspond to mean ± standard deviation for accuracy, balanced accuracy, ROC-AUC, F1-score, precision, and recall, computed after collapsing repeated raw outputs at the subject-session-pipeline level and averaging across sessions when applicable. The table also reports the number of subjects, number of pipelines, and number of subject-level observations contributing to each estimate.

| Dataset | Method | Family | Band | Subjects (n) | Pipelines (n) | Subject-pipeline observations (n) | Accuracy | Balanced accuracy | ROC-AUC | F1-score | Precision | Recall |
|---|---|---|---|---|---|---|---|---|---|---|---|---|
| Cho2017 | con_instantaneous+tgsp | connectivity_tgsp | 8-15 | 52 | 88 | 4576 | 0.630 ± 0.115 | 0.629 ± 0.116 | 0.629 ± 0.115 | 0.629 ± 0.116 | 0.629 ± 0.116 | 0.629 ± 0.115 |
| Cho2017 | cov_tgsp | cov_tgsp | 8-15 | 52 | 80 | 4160 | 0.625 ± 0.121 | 0.624 ± 0.121 | 0.625 ± 0.122 | 0.624 ± 0.122 | 0.625 ± 0.121 | 0.625 ± 0.121 |
| Cho2017 | csp_2 | csp | 8-15 | 52 | 88 | 4576 | 0.667 ± 0.156 | 0.666 ± 0.157 | 0.666 ± 0.157 | 0.667 ± 0.157 | 0.667 ± 0.157 | 0.667 ± 0.156 |
| Cho2017 | hfd | nonlinear | 8-15 | 52 | 72 | 3744 | 0.533 ± 0.065 | 0.533 ± 0.064 | 0.532 ± 0.065 | 0.533 ± 0.064 | 0.533 ± 0.065 | 0.532 ± 0.065 |
| Cho2017 | hjorth | nonlinear | 8-15 | 52 | 88 | 4576 | 0.550 ± 0.083 | 0.550 ± 0.083 | 0.550 ± 0.083 | 0.551 ± 0.083 | 0.550 ± 0.082 | 0.550 ± 0.083 |
| Cho2017 | svd_entropy | nonlinear | 8-15 | 52 | 88 | 4576 | 0.555 ± 0.089 | 0.555 ± 0.089 | 0.555 ± 0.088 | 0.555 ± 0.089 | 0.554 ± 0.088 | 0.555 ± 0.088 |
| Cho2017 | con_instantaneous+tgsp | connectivity_tgsp | 8-30 | 52 | 88 | 4576 | 0.604 ± 0.118 | 0.604 ± 0.119 | 0.604 ± 0.119 | 0.604 ± 0.119 | 0.604 ± 0.119 | 0.604 ± 0.119 |
| Cho2017 | cov_tgsp | cov_tgsp | 8-30 | 52 | 16 | 832 | 0.712 ± 0.140 | 0.709 ± 0.140 | 0.710 ± 0.138 | 0.709 ± 0.139 | 0.709 ± 0.138 | 0.711 ± 0.139 |
| Cho2017 | csp_2 | csp | 8-30 | 52 | 88 | 4576 | 0.661 ± 0.158 | 0.660 ± 0.158 | 0.661 ± 0.157 | 0.660 ± 0.158 | 0.660 ± 0.158 | 0.661 ± 0.157 |
| Cho2017 | hfd | nonlinear | 8-30 | 52 | 24 | 1248 | 0.586 ± 0.092 | 0.586 ± 0.092 | 0.586 ± 0.091 | 0.586 ± 0.092 | 0.585 ± 0.092 | 0.586 ± 0.091 |
| Cho2017 | hjorth | nonlinear | 8-30 | 52 | 56 | 2912 | 0.563 ± 0.094 | 0.564 ± 0.094 | 0.564 ± 0.093 | 0.563 ± 0.093 | 0.564 ± 0.094 | 0.563 ± 0.094 |
| Cho2017 | svd_entropy | nonlinear | 8-30 | 52 | 88 | 4576 | 0.582 ± 0.093 | 0.582 ± 0.092 | 0.581 ± 0.092 | 0.582 ± 0.093 | 0.582 ± 0.093 | 0.581 ± 0.092 |
| PhysionetMotorImagery | con_instantaneous+tgsp | connectivity_tgsp | 8-15 | 109 | 88 | 9592 | 0.520 ± 0.121 | 0.521 ± 0.121 | 0.521 ± 0.121 | 0.520 ± 0.120 | 0.522 ± 0.120 | 0.520 ± 0.120 |
| PhysionetMotorImagery | cov_tgsp | cov_tgsp | 8-15 | 109 | 88 | 9592 | 0.583 ± 0.133 | 0.583 ± 0.132 | 0.583 ± 0.132 | 0.584 ± 0.132 | 0.584 ± 0.132 | 0.584 ± 0.132 |
| PhysionetMotorImagery | csp_2 | csp | 8-15 | 109 | 88 | 9592 | 0.596 ± 0.172 | 0.596 ± 0.172 | 0.596 ± 0.173 | 0.597 ± 0.172 | 0.597 ± 0.172 | 0.597 ± 0.172 |
| PhysionetMotorImagery | hfd | nonlinear | 8-15 | 109 | 88 | 9592 | 0.534 ± 0.119 | 0.534 ± 0.119 | 0.535 ± 0.119 | 0.535 ± 0.119 | 0.534 ± 0.119 | 0.534 ± 0.119 |
| PhysionetMotorImagery | hjorth | nonlinear | 8-15 | 109 | 88 | 9592 | 0.527 ± 0.113 | 0.527 ± 0.112 | 0.527 ± 0.112 | 0.528 ± 0.113 | 0.528 ± 0.112 | 0.527 ± 0.114 |
| PhysionetMotorImagery | svd_entropy | nonlinear | 8-15 | 109 | 88 | 9592 | 0.534 ± 0.125 | 0.534 ± 0.126 | 0.535 ± 0.126 | 0.534 ± 0.126 | 0.535 ± 0.126 | 0.535 ± 0.125 |
| PhysionetMotorImagery | con_instantaneous+tgsp | connectivity_tgsp | 8-30 | 109 | 40 | 4360 | 0.495 ± 0.118 | 0.494 ± 0.119 | 0.495 ± 0.119 | 0.495 ± 0.118 | 0.495 ± 0.116 | 0.494 ± 0.119 |
| PhysionetMotorImagery | cov_tgsp | cov_tgsp | 8-30 | 109 | 80 | 8720 | 0.605 ± 0.174 | 0.606 ± 0.173 | 0.606 ± 0.173 | 0.606 ± 0.174 | 0.606 ± 0.174 | 0.605 ± 0.173 |
| PhysionetMotorImagery | csp_2 | csp | 8-30 | 109 | 88 | 9592 | 0.604 ± 0.178 | 0.605 ± 0.179 | 0.603 ± 0.179 | 0.605 ± 0.178 | 0.605 ± 0.178 | 0.604 ± 0.179 |
| PhysionetMotorImagery | hfd | nonlinear | 8-30 | 109 | 88 | 9592 | 0.528 ± 0.130 | 0.529 ± 0.129 | 0.530 ± 0.129 | 0.529 ± 0.130 | 0.529 ± 0.130 | 0.529 ± 0.130 |
| PhysionetMotorImagery | hjorth | nonlinear | 8-30 | 109 | 88 | 9592 | 0.532 ± 0.126 | 0.531 ± 0.126 | 0.531 ± 0.126 | 0.533 ± 0.126 | 0.532 ± 0.126 | 0.533 ± 0.126 |
| PhysionetMotorImagery | svd_entropy | nonlinear | 8-30 | 109 | 88 | 9592 | 0.530 ± 0.126 | 0.530 ± 0.126 | 0.530 ± 0.126 | 0.529 ± 0.125 | 0.530 ± 0.126 | 0.531 ± 0.126 |
| Zhou2016 | con_instantaneous+tgsp | connectivity_tgsp | 8-15 | 4 | 88 | 352 | 0.646 ± 0.115 | 0.647 ± 0.113 | 0.647 ± 0.115 | 0.647 ± 0.115 | 0.646 ± 0.113 | 0.646 ± 0.115 |
| Zhou2016 | cov_tgsp | cov_tgsp | 8-15 | 4 | 88 | 352 | 0.752 ± 0.112 | 0.752 ± 0.114 | 0.753 ± 0.112 | 0.753 ± 0.113 | 0.750 ± 0.113 | 0.753 ± 0.113 |
| Zhou2016 | csp_2 | csp | 8-15 | 4 | 88 | 352 | 0.832 ± 0.121 | 0.831 ± 0.121 | 0.831 ± 0.122 | 0.829 ± 0.121 | 0.830 ± 0.121 | 0.831 ± 0.122 |
| Zhou2016 | hfd | nonlinear | 8-15 | 4 | 88 | 352 | 0.615 ± 0.115 | 0.615 ± 0.114 | 0.615 ± 0.115 | 0.615 ± 0.115 | 0.615 ± 0.115 | 0.614 ± 0.115 |
| Zhou2016 | hjorth | nonlinear | 8-15 | 4 | 88 | 352 | 0.631 ± 0.100 | 0.630 ± 0.100 | 0.629 ± 0.098 | 0.630 ± 0.100 | 0.629 ± 0.099 | 0.629 ± 0.099 |
| Zhou2016 | svd_entropy | nonlinear | 8-15 | 4 | 88 | 352 | 0.598 ± 0.095 | 0.598 ± 0.094 | 0.598 ± 0.093 | 0.597 ± 0.093 | 0.597 ± 0.092 | 0.598 ± 0.093 |
| Zhou2016 | con_instantaneous+tgsp | connectivity_tgsp | 8-30 | 4 | 80 | 320 | 0.606 ± 0.088 | 0.606 ± 0.089 | 0.605 ± 0.089 | 0.607 ± 0.089 | 0.607 ± 0.091 | 0.605 ± 0.090 |
| Zhou2016 | cov_tgsp | cov_tgsp | 8-30 | 4 | 88 | 352 | 0.756 ± 0.115 | 0.756 ± 0.115 | 0.756 ± 0.115 | 0.755 ± 0.115 | 0.756 ± 0.114 | 0.756 ± 0.115 |
| Zhou2016 | csp_2 | csp | 8-30 | 4 | 88 | 352 | 0.818 ± 0.138 | 0.817 ± 0.137 | 0.817 ± 0.138 | 0.818 ± 0.139 | 0.817 ± 0.136 | 0.815 ± 0.136 |
| Zhou2016 | hfd | nonlinear | 8-30 | 4 | 88 | 352 | 0.647 ± 0.087 | 0.647 ± 0.086 | 0.645 ± 0.086 | 0.645 ± 0.085 | 0.645 ± 0.086 | 0.646 ± 0.086 |
| Zhou2016 | hjorth | nonlinear | 8-30 | 4 | 88 | 352 | 0.593 ± 0.085 | 0.594 ± 0.085 | 0.593 ± 0.085 | 0.595 ± 0.083 | 0.591 ± 0.083 | 0.595 ± 0.086 |
| Zhou2016 | svd_entropy | nonlinear | 8-30 | 4 | 88 | 352 | 0.645 ± 0.113 | 0.644 ± 0.112 | 0.645 ± 0.112 | 0.644 ± 0.112 | 0.644 ± 0.111 | 0.644 ± 0.112 |

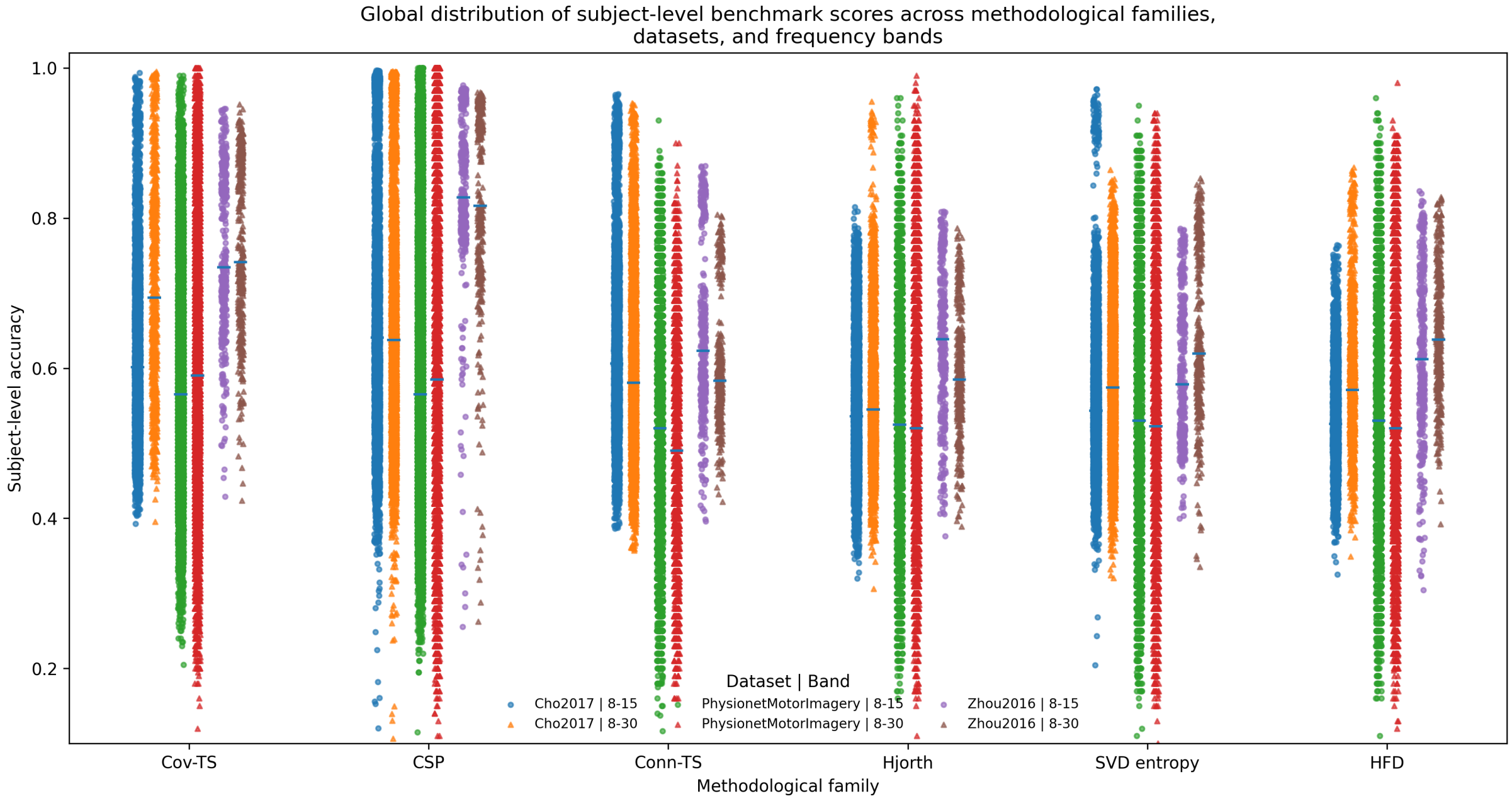

**Figure 1.** Global distribution of subject-level benchmark scores across methodological families, datasets, and frequency bands. Each point represents an aggregated subject-level pipeline score after collapsing repeated raw outputs at the subject-session-pipeline level and averaging across sessions when applicable. Methods are displayed by family on the x-axis, while marker position and grouping allow visual comparison across datasets and frequency bands (8-15 Hz and 8-30 Hz). Small horizontal ticks indicate the median score within each dataset-band-method combination.

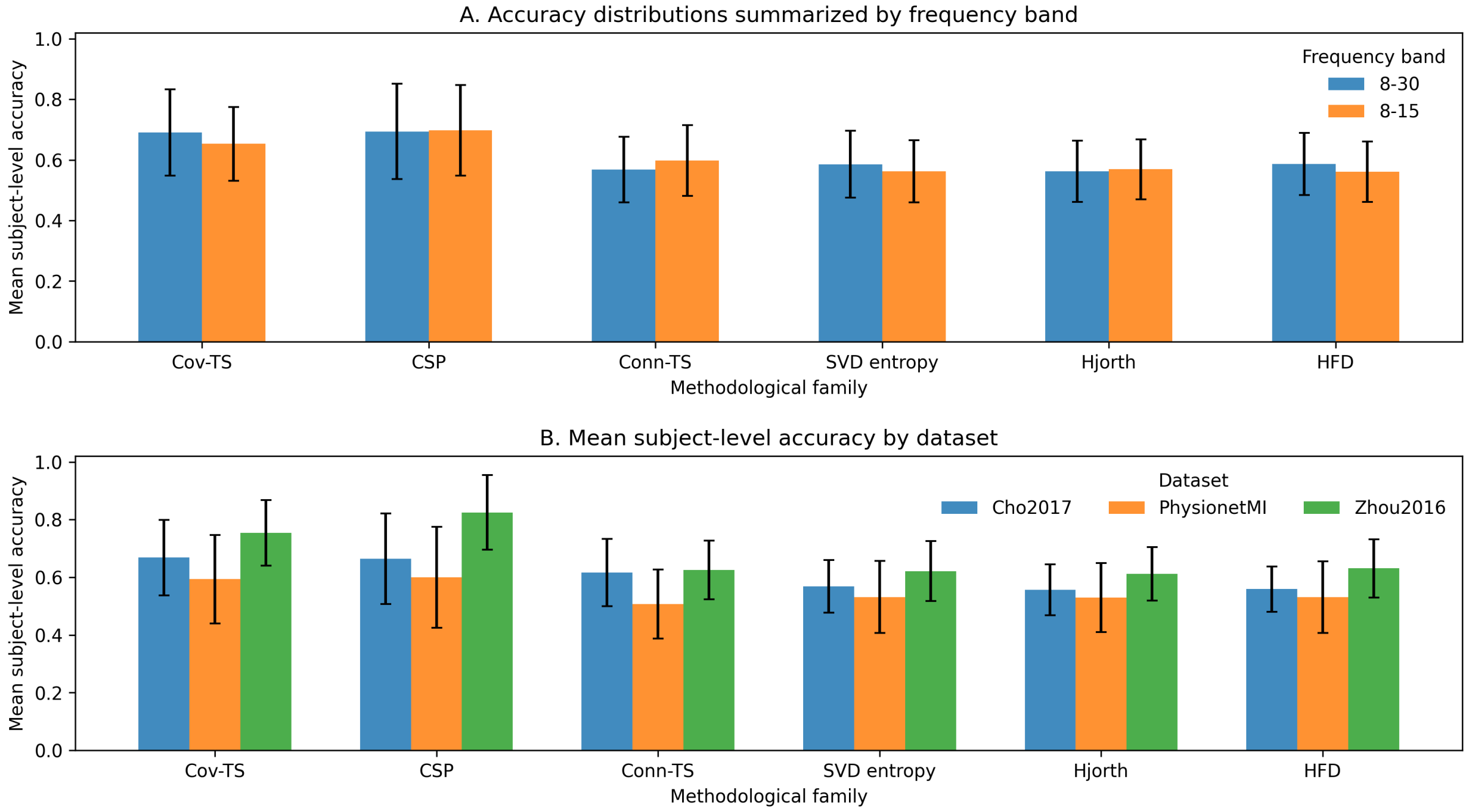

**Figure 2.** Comparative performance of methodological families across frequency bands and datasets. Panel A shows mean subject-level accuracy for each methodological family in the 8-15 Hz and 8-30 Hz bands after averaging across datasets. Panel B shows mean subject-level accuracy for the same families in Cho2017, PhysionetMI, and Zhou2016 after averaging across frequency bands. Error bars denote standard deviations derived from subject-level aggregated benchmark scores. This figure complements Figure 1 by separating the effects of frequency band and dataset on the relative ranking of methodological families.

## Heterogeneity of optimal configurations across datasets and subjects

The benchmark-level results identified cov-tgsp and CSP as the leading methodological families overall. However, these aggregate rankings did not translate into a single universally optimal configuration at the subject level. As summarized in **Table 2** and illustrated in **Figure 3**, the identity of the best-performing configuration varied markedly across datasets and individuals, not only in terms of feature family, but also with respect to the complete pipeline, preprocessing strategy, classifier, and frequency band.

A first important result is that the best-performing feature family was not stable across subjects. On Cho2017, Cov-TS was the most frequent winning family, accounting for 26 of 52 subjects, followed by CSP with 18 winning subjects (**Figure 3A-B**; **Table 2**). The remaining four families each won only a small number of subjects, but their presence confirms that the leading benchmark-level families did not capture all individual optima. On PhysionetMI, the subject-level distribution was substantially flatter: Cov-TS won 29 subjects and CSP won 28, while the remaining winners were distributed across Hjorth (19), Conn-TS (12), SVD entropy (11), and HFD (10). Thus, unlike Cho2017, PhysionetMI did not exhibit a clear dominance of a single family at the individual level. On Zhou2016, all 4 subjects were won by CSP, but this apparent family-level uniformity should be interpreted cautiously given the very small cohort size.

The same pattern became even more striking when the analysis moved from feature families to complete pipelines. On Cho2017, the 52 subjects were associated with 42 distinct winning pipelines, indicating that only limited repetition of the exact same end-to-end configuration occurred across individuals (**Table 2**). On PhysionetMI, the heterogeneity was even stronger, with 93 distinct winning pipelines across 109 subjects. Even on Zhou2016, where all winners belonged to the same family, the 4 subjects were still associated with 4 distinct winning pipelines. These results show that family-level regularities mask a much more fragmented configuration landscape once the full decoding pipeline is considered.

This fragmentation was also visible when the winning pipelines were decomposed into their classifiers (**Figure 3D**; **Table 2**). On Cho2017, the most frequent winning classifier was logistic regression, but it accounted for only 10 of 52 subjects, well below the total cohort size. The remaining winning subjects were distributed across a wide range of models, including MLP-3 (7), linear SVM (5), MLP-11 (5), and several additional MLP variants, indicating that no single classifier dominated the dataset. On PhysionetMI, classifier heterogeneity was even more pronounced: logistic regression and random forest each won 12 subjects, followed by linear SVM (8), LDA (7), MLP-4 (7), MLP-5 (7), and MLP-7 (7), with many other classifiers still contributing non-negligible numbers of wins. Thus, even when feature families showed some regularity at the aggregate level, the downstream model remained highly unstable across individual optima. On Zhou2016, the four subject-level winners were associated with four distinct classifiers, again indicating that a single best classifier did not emerge even in the smallest dataset.

A similar pattern was observed for preprocessing strategies (**Figure 3C**; **Table 2**). On Cho2017, MinMax scaling was the most frequent winning scaler, accounting for 30 subjects, followed by StandardScaler (9), RobustScaler (7), and L2 normalization (6). In contrast, PhysionetMI showed a much flatter distribution, with StandardScaler (37) and MinMax (36) nearly tied, followed by RobustScaler (23) and L2 normalization (13). This indicates that scaler choice modulated the final optimum without converging toward a uniquely dominant preprocessing strategy. On Zhou2016, each of the four scalers won one subject, further illustrating the instability of the full optimal configuration.

The frequency band also contributed to this heterogeneity. On Cho2017, the best-performing pipelines were more often associated with the broader 8-30 Hz range (29 subjects) than with 8-15 Hz (23 subjects). The preference for 8-30 Hz was even more pronounced on PhysionetMI, where it accounted for 70 winning subjects versus 39 for 8-15 Hz. In contrast, Zhou2016 showed the opposite tendency, with 8-15 Hz winning 3 of 4 subjects (**Table 2**). Thus, band preference was not universal and depended strongly on dataset context.

A final indicator of configuration ambiguity was the prevalence of tie cases before tie-breaking. Only 1 tie case was observed on Cho2017, whereas PhysionetMI showed 21, by far the highest number among the three datasets (**Table 2**). This suggests that on PhysionetMI, multiple configurations often reached very similar best-subject performance, reinforcing the impression of a flatter and more ambiguous optimization landscape. Zhou2016 showed 1 tie case, but given the very small cohort, this should not be overinterpreted.

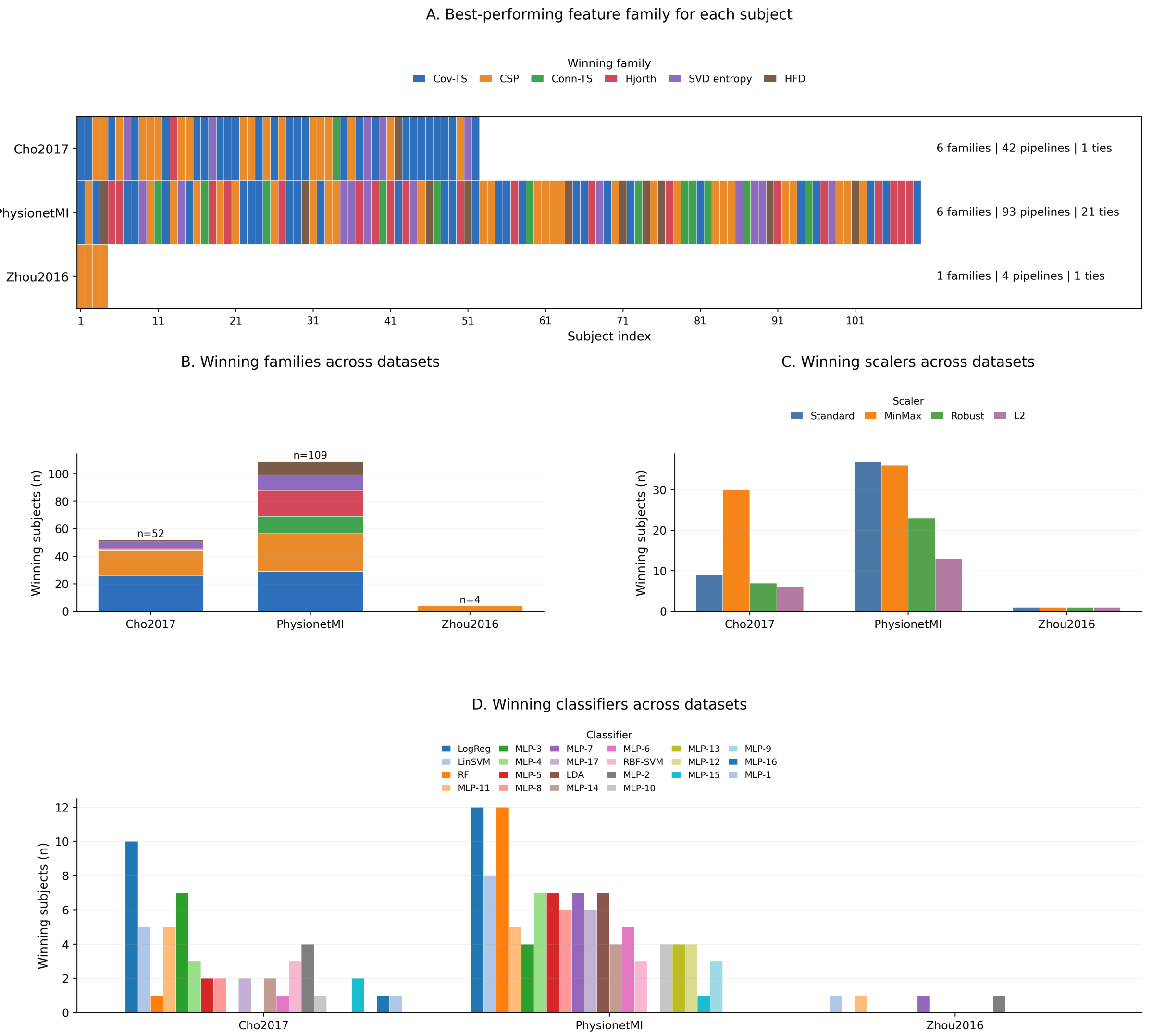

**Figure 3.** Subject- and dataset-level heterogeneity of optimal decoding configurations. Panel A displays the best-performing feature family for each subject across Cho2017, PhysionetMI, and Zhou2016. Panel B reports the distribution of winning feature families across datasets. Panel C reports the distribution of winning preprocessing scalers across datasets. Panel D reports the distribution of winning classifiers across datasets. Winning configurations were defined from subject-level aggregated benchmark scores.

**Table 2.** Summary of subject-level winning configuration heterogeneity across datasets. The table summarizes the heterogeneity of subject-level optimal decoding configurations across datasets. For each dataset, the table reports the number of distinct winning feature families and winning full pipelines, together with the most frequent winning family, classifier, scaler, and frequency band. Tie cases before tie-breaking are also reported to quantify the frequency of near-equivalent optimal configurations.

| Dataset | Subjects (n) | Distinct winning families (n) | Most frequent winning family | Winning family count | Distinct winning pipelines (n) | Most frequent winning classifier | Winning classifier count | Most frequent winning scaler | Winning scaler count | Most frequent winning band | Winning band count | Tie cases before tie-break (n) |
|---|---|---|---|---|---|---|---|---|---|---|---|---|
| Cho2017 | 52 | 6 | Cov-TS | 26 | 42 | Logistic regression | 10 | MinMax scaling | 30 | 8-30 Hz | 29 | 1 |
| PhysionetMI | 109 | 6 | Cov-TS | 29 | 93 | Logistic regression<br>Random forest | 12 | Standard scaling | 37 | 8-30 Hz | 70 | 21 |
| Zhou2016 | 4 | 1 | CSP | 4 | 4 | One per winning subject | 1 | One per winning subject | 1 | 8-15 Hz | 3 | 1 |

## Secondary metrics and robustness checks

The benchmark-level hierarchy described above was primarily established from subject-level accuracy, but the same comparisons were also examined using additional evaluation metrics in order to assess the robustness of the main findings. Overall, the leading methodological families identified on accuracy remained the strongest when evaluated with balanced accuracy, indicating that the observed ranking was not driven solely by dataset-specific class imbalance or by a favorable response to the dominant class.

The strongest consistency was observed for the two leading families, cov-tgsp and CSP. On Cho2017, the best-performing family-level configuration, cov-tgsp in 8-30 Hz, achieved a mean balanced accuracy of 0.709 ± 0.140, closely matching its mean accuracy (0.712 ± 0.140). The same pattern held for CSP, whose balanced accuracy remained very close to the corresponding accuracy values in both 8-15 Hz and 8-30 Hz. Thus, on Cho2017, the main hierarchy observed on accuracy remained essentially unchanged when assessed using balanced accuracy.

A similar robustness was observed on PhysionetMI, although the absolute performance level was lower and the separation between the leading families was smaller. Cov-tgsp and CSP remained the two strongest families under balanced accuracy, just as they did for accuracy, while the nonlinear feature families again formed a lower and relatively compressed group. This confirms that the flatter ranking observed on PhysionetMI was not specific to a single metric, but reflected a more general reduction in separability between competing approaches in the most heterogeneous dataset of the benchmark.

On Zhou2016, CSP also preserved its leading position under balanced accuracy, followed by cov-tgsp, with the remaining families occupying clearly lower ranks. Although the very small cohort size limits strong conclusions on this dataset, the preservation of the same family-level ordering under both accuracy and balanced accuracy again supports the robustness of the main benchmark trends.

The same general pattern was observed for the other secondary metrics. Across datasets, the leading families identified from accuracy remained among the strongest when evaluated with ROC-AUC, F1-score, precision, and recall, whereas coherence-based tangent-space representations and nonlinear descriptors remained globally less competitive. While the exact magnitude of the differences varied across metrics and datasets, these additional evaluations did not reveal a qualitatively different benchmark structure.

Taken together, these robustness checks indicate that the main conclusions of the benchmark were not critically dependent on the use of accuracy alone. Instead, the overall hierarchy remained stable across multiple evaluation criteria, with cov-tgsp and CSP consistently defining the leading group at the family level. Detailed benchmark results for all secondary metrics, including balanced accuracy, ROC-AUC, F1-score, precision, and recall, are provided in the supplementary material (**supplementray_material_benchmark**).

## Statistical comparison of the main methodological families

To complement the descriptive benchmark analyses, we performed a statistical comparison of the main methodological families at the subject level. For each dataset, and for each subject, the score retained for a given family corresponded to the best-performing pipeline within that family, based on accuracy. This yielded paired subject-level observations across the six methodological families (Cov-TS, CSP, Conn-TS, Hjorth, SVD entropy, and HFD), allowing direct within-subject comparison. Global differences were first assessed using a Friedman test, followed, when appropriate, by pairwise Wilcoxon signed-rank tests with Holm correction for multiple comparisons.

The Friedman test indicated a significant overall family effect in all three datasets. On Cho2017, the effect was strong ($\chi^2 = 99.04$, $p = 8.42 \times 10^{-20}$), consistent with the clear hierarchy observed in the descriptive analyses. On PhysionetMI, the overall family effect remained significant but smaller ($\chi^2 = 22.25$, $p = 4.69 \times 10^{-4}$), reflecting the flatter performance landscape observed in the largest and most heterogeneous dataset. On Zhou2016, the Friedman test was also significant ($\chi^2 = 14.43$, $p = 0.0131$), although the very small number of subjects ($n = 4$) requires cautious interpretation.

Post-hoc pairwise comparisons showed that the strongest statistical separation was observed on Cho2017. After Holm correction, Cov-TS significantly outperformed HFD ($p = 1.57 \times 10^{-8}$), Conn-TS ($p = 1.50 \times 10^{-7}$), Hjorth ($p = 2.25 \times 10^{-7}$), and SVD entropy ($p = 7.78 \times 10^{-7}$). CSP also significantly outperformed HFD ($p = 1.58 \times 10^{-6}$), Hjorth ($p = 2.87 \times 10^{-6}$), SVD entropy ($p = 4.44 \times 10^{-6}$), and Conn-TS ($p = 1.15 \times 10^{-5}$). No significant difference was observed between Cov-TS and CSP after correction, indicating that although Cov-TS was numerically strongest on Cho2017, the two leading families were not statistically separated from each other.

On PhysionetMI, the post-hoc structure was more compressed. Cov-TS significantly outperformed Conn-TS ($p = 9.00 \times 10^{-5}$), SVD entropy ($p = 0.0130$), and HFD ($p = 0.0302$), while CSP significantly outperformed Conn-TS ($p = 0.00133$). In contrast, Cov-TS and CSP were again not significantly different from each other, consistent with their near-equivalence in the descriptive results. Likewise, Hjorth did not differ significantly from the leading families after correction. Thus, the statistical analysis confirmed that PhysionetMI exhibited weaker family-level separation than Cho2017.

On Zhou2016, no pairwise comparison remained significant after Holm correction despite the significant overall Friedman result. Given the very small cohort size, this should not be interpreted as evidence of equivalence between families, but rather as a limitation of statistical power. The descriptive analyses nevertheless remained consistent with CSP as the strongest family and Cov-TS as the second-best family on this dataset.

To further assess whether fine-grained differences persisted within the two leading families, we also compared the top two subject-level pipelines within Cov-TS and CSP in each dataset using paired Wilcoxon signed-rank tests on subject-level accuracy. These within-family comparisons did not reveal significant differences between the leading pipelines. On Cho2017, the best two Cov-TS pipelines (8-30 Hz + MinMax + linear SVM and 8-30 Hz + MinMax + elastic-net logistic regression) were statistically indistinguishable ($p = 0.699$), as were the top two CSP pipelines ($p = 0.554$). The same pattern held on PhysionetMI, where the top two Cov-TS pipelines did not differ significantly ($p = 0.082$), and the top two CSP pipelines were likewise indistinguishable ($p = 0.884$). On Zhou2016, no within-family comparison reached significance for either Cov-TS ($p = 0.875$) or CSP ($p = 1.000$), again with the caveat of limited statistical power. These results suggest that, within the two dominant families, several top-performing pipelines were effectively interchangeable at the subject level.

Taken together, these statistical comparisons refine the main conclusions of the benchmark in three ways. First, they confirm that the benchmark-level hierarchy was not purely descriptive, but reflected statistically detectable differences between methodological families, particularly on Cho2017. Second, they show that the two leading families, Cov-TS and CSP, were not significantly separated from each other in the larger datasets, despite consistently occupying the top positions. Third, they indicate that even within these dominant families, the best individual pipelines were often statistically indistinguishable. Thus, the benchmark supports the existence of leading methodological families, but not that of a single universally superior pipeline.

## Subject-specific oracle recovery as a function of portfolio size

Given the substantial subject-level heterogeneity described above, we next examined whether compact portfolios of pipelines could recover a larger fraction of the subject-specific oracle than a single fixed global pipeline (**Figure 4**). For interpretability, $K = 1$ was defined as the best single global pipeline, whereas $K > 1$ corresponded to optimized portfolios selected with the topk_mean portfolio-selection heuristic, which was retained for the primary analysis because it combined strong empirical performance with a simple and interpretable selection rule.

Portfolio size had a clear but dataset-dependent effect. In Cho2017, oracle retention increased progressively with $K$, from 0.942 to 0.965 in the 8–15 Hz band and from 0.936 to 0.965 in the 8–30 Hz band between $K = 1$ and $K = 12$. In absolute terms, this corresponded to gains of +0.019 and +0.024 balanced-accuracy points, respectively, relative to the best single global pipeline. Most of the improvement occurred at small-to-moderate portfolio sizes, after which the curves approached a plateau.

The effect was substantially larger in PhysionetMotorImagery. Oracle retention rose from 0.818 to 0.900 in the 8–15 Hz band and from 0.804 to 0.902 in the 8–30 Hz band when increasing portfolio size from $K = 1$ to $K = 12$. This translated into absolute gains of +0.069 and +0.082 balanced-accuracy points, respectively. Unlike Cho2017, PhysionetMotorImagery showed continued improvement even at larger $K$, indicating a broader and more complex subject-dependent optimization space.

In contrast, Zhou2016 showed no practical benefit from portfolio optimization. Oracle retention was already near ceiling at $K = 1$, and larger portfolios produced slightly lower values, reaching 0.984 in the 8–15 Hz band and 0.988 in the 8–30 Hz band at $K = 12$. Accordingly, the absolute gain relative to the single best global pipeline was slightly negative. Given the very small number of available subjects and the already high baseline performance, Zhou2016 is more consistent with a saturated regime than with meaningful portfolio complementarity.

An additional result was that the diversity exploited by portfolio selection arose predominantly within the globally dominant feature family rather than from broad combinations across many families. In Cho2017, the selected topk_mean portfolios were entirely composed of cov_tgsp pipelines across all tested portfolio sizes and both frequency bands. PhysionetMotorImagery showed the same overall pattern, with portfolios almost exclusively dominated by cov_tgsp variants. Conversely, all selected portfolios in Zhou2016 were CSP-based. Thus, the main benefit of portfolio optimization did not stem from mixing heterogeneous feature families, but rather from exploiting complementary classifier or scaling variants within the family that already dominated global performance.

Taken together, these results show that the value of portfolio optimization is strongly constrained by the degree of subject-level heterogeneity. When heterogeneity is high, as in PhysionetMotorImagery and to a lesser extent Cho2017, compact portfolios recover a substantially larger fraction of the subject-specific oracle than a single fixed global pipeline. When heterogeneity is limited and global performance is already near ceiling, as in Zhou2016, the room for improvement becomes negligible.

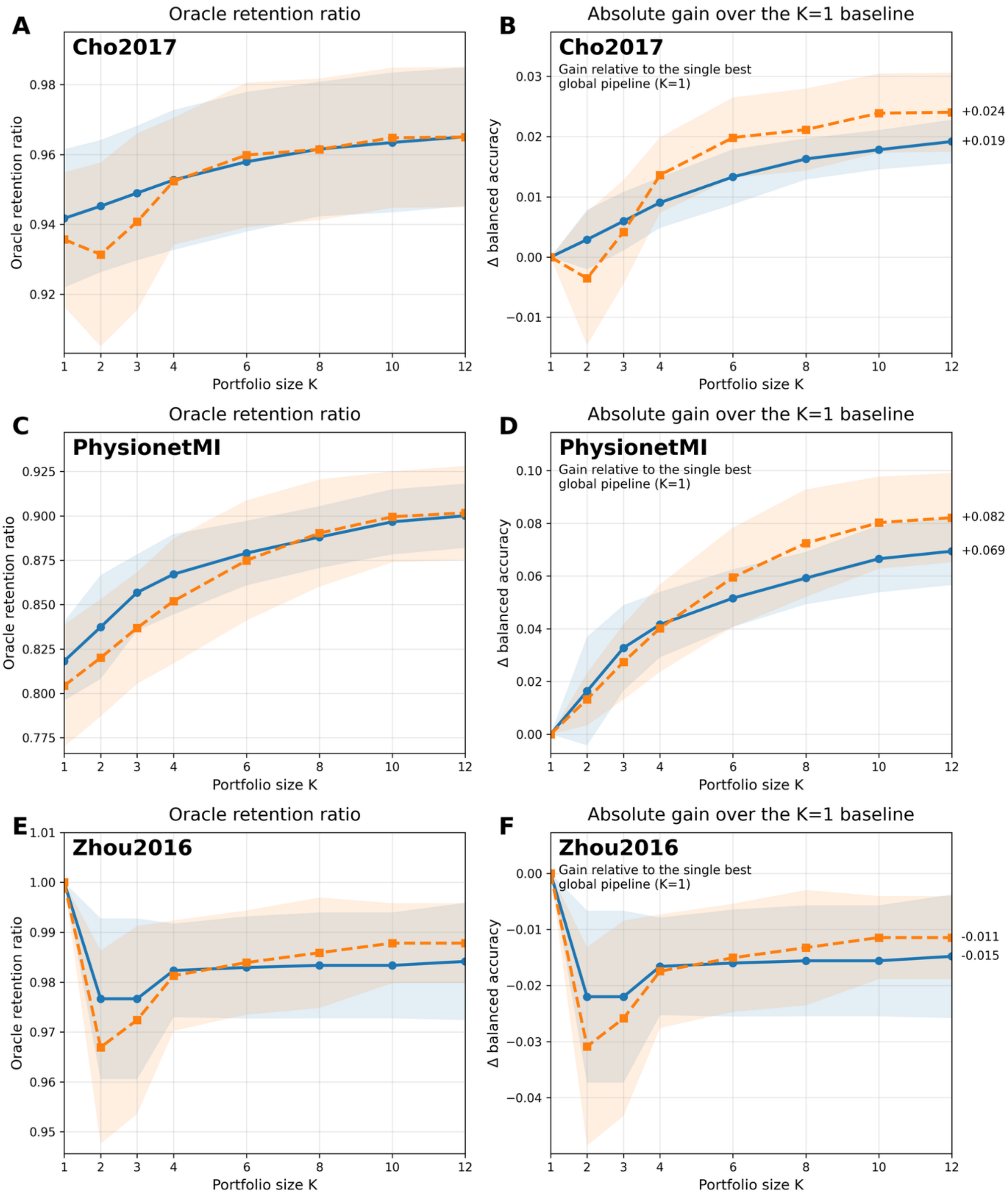

**Figure 4.** Portfolio optimization performance across datasets and portfolio sizes. Multi-panel summary of portfolio optimization results using the topk_mean portfolio-selection heuristic. Rows correspond to Cho2017, PhysionetMotorImagery, and Zhou2016. Left-column panels show the oracle retention ratioas a function of portfolio size $K$, where $K = 1$ denotes the fixed best single global pipeline and larger $K$ values denote optimized portfolios. Right-column panels show the corresponding absolute gain in balanced accuracy relative to the $K = 1$ baseline. Curves are shown separately for the 8–15 Hz and 8–30 Hz bands, and shaded regions indicate fold-wise standard deviation. Cho2017 and PhysionetMotorImagery exhibited

progressive improvement with increasing $K$, with the largest gains observed in PhysionetMotorImagery, whereas Zhou2016 showed near-saturated performance already at

## Robustness across portfolio-construction strategies

To determine whether these conclusions depended on a specific portfolio-selection procedure, we compared seven strategies: beam, greedy, greedy_diverse, regret_greedy, submodular_coverage, family_constrained, and topk_mean (**Figure 5**). Across all 42 evaluated dataset-band-$K$ combinations, the average oracle-retention ratio was highest for topk_mean at 0.936, followed by regret_greedy at 0.932, greedy at 0.930, beam at 0.929, greedy_diverse at 0.925, submodular_coverage at 0.918, and family_constrained at 0.904. The corresponding mean ranks across the six dataset-by-band configurations were 1.50 for topk_mean, 1.67 for regret_greedy, 3.00 for greedy, 4.17 for beam, 5.00 for greedy_diverse, and 6.33 for both submodular_coverage and family_constrained.

At the dataset level, topk_mean ranked first in four of the six dataset-by-band configurations: Cho2017 8–15 Hz (0.956), Cho2017 8–30 Hz (0.954), Zhou2016 8–15 Hz (0.981), and Zhou2016 8–30 Hz (0.981). Regret_greedy ranked first in the two PhysionetMotorImagery settings, with mean oracle-retention ratios of 0.881 in 8–15 Hz and 0.872 in 8–30 Hz, compared with 0.875 and 0.868 for topk_mean, respectively. Thus, although topk_mean was not uniformly best in every configuration, it provided the strongest average performance overall and the best average rank.

The same general conclusion held across portfolio sizes. All strategies improved progressively with increasing $K$, confirming that the main effect was not specific to a single selection rule. The mean oracle-retention ratio of topk_mean increased from 0.913 at $K = 2$ to 0.951 at $K = 12$, while regret_greedy increased from 0.907 to 0.949, greedyfrom 0.907 to 0.947, and beam from 0.906 to 0.946. Greedy_diverse also improved steadily, from 0.906 to 0.941. The weakest early performance was observed for family_constrained, which started at 0.835 for $K = 2$, but its performance increased markedly with larger portfolios, reaching 0.938 at $K = 12$. By contrast, submodular_coverage remained consistently below the best-performing strategies, improving from 0.906 at $K = 2$ to 0.929 at $K = 12$.

Importantly, the differences between the strongest methods were modest relative to the larger effects induced by dataset heterogeneity and portfolio size. For example, in Cho2017 8–15 Hz, topk_mean and regret_greedy differed by only 0.0002 in mean oracle retention. In PhysionetMotorImagery 8–15 Hz, the gap between regret_greedy and topk_meanwas 0.0057, and in PhysionetMotorImagery 8–30 Hz the gap between regret_greedy and greedy was 0.0027. These small differences indicate that the main finding of the study does not depend on a single algorithmic choice.

The weakest results were obtained overall with family_constrained, which achieved the lowest average oracle retention and ranked last in Cho2017 8–15 Hz, Cho2017 8–30 Hz, Zhou2016 8–15 Hz, and Zhou2016 8–30 Hz. This observation is consistent with the composition analysis of selected portfolios: explicitly forcing broad family diversity did not provide a systematic advantage, because most useful diversity arose within the already dominant family rather than across unrelated feature families.

Overall, these comparisons show that portfolio optimization is robust to the specific selection strategy used. Although topk_mean offered the best global trade-off and was therefore retained for the main analysis, regret_greedy, greedy, and beam produced closely related results across datasets and across portfolio sizes. This convergence supports the interpretation that the observed gains reflect a genuine structural property of the benchmark, namely the presence of subject-specific variability that can be partially captured by compact portfolios.

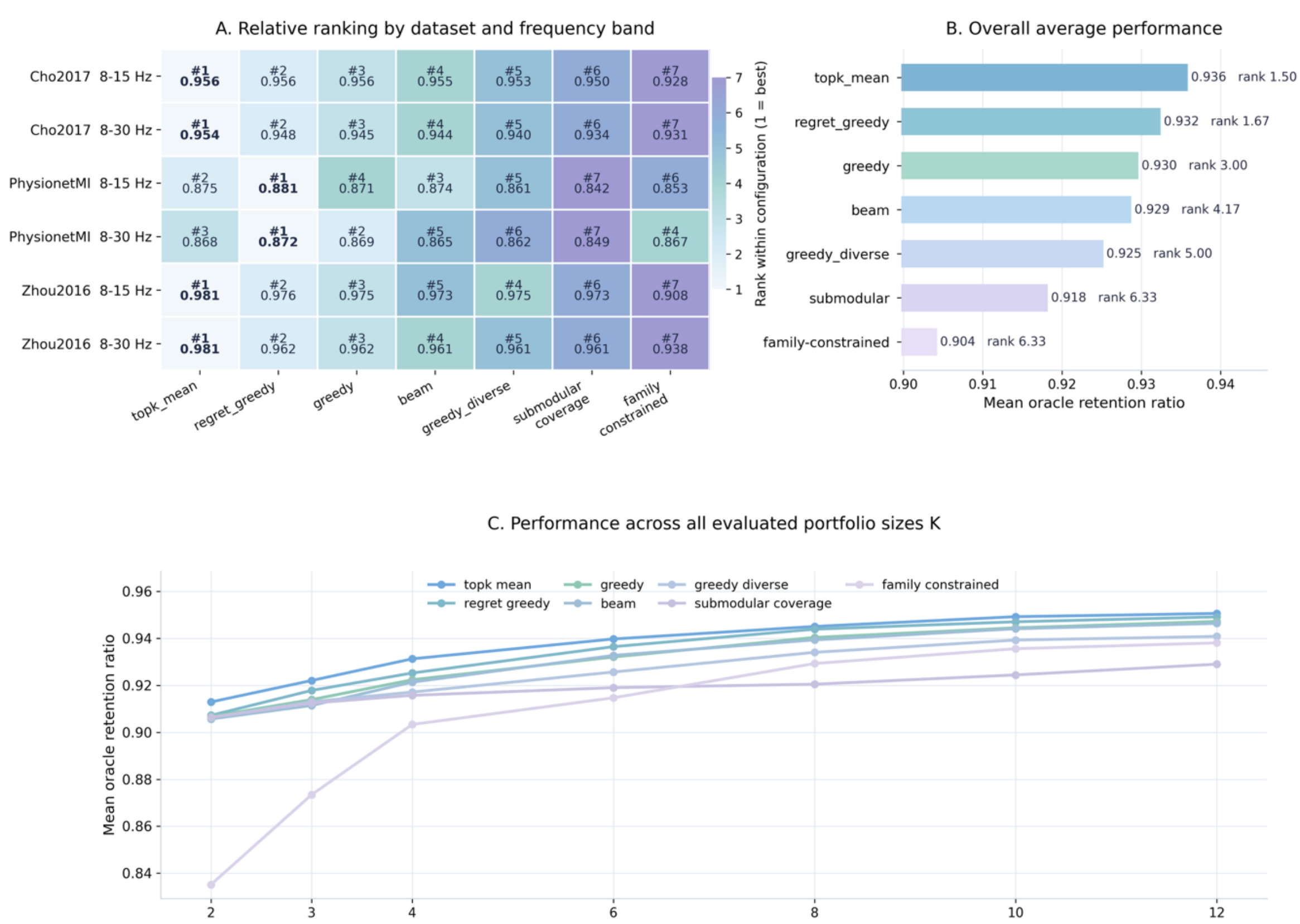

**Figure 5.** Comparison of portfolio-construction strategies. Panel A shows the rank of each strategy within each dataset-by-band configuration, based on mean oracle-retention ratio averaged across the evaluated portfolio sizes. Cell labels report both rank and mean oracle-retention value. Panel B summarizes the overall mean oracle-retention ratio of each strategy across all dataset-by-band configurations. Panel C shows how mean oracle-retention performance evolved with portfolio size *K* for each strategy, averaged across dataset and frequency-band settings. Across analyses, topk_mean provided the strongest overall trade-off, while regret_greedy remained highly competitive and family_constrained underperformed especially at small portfolio sizes.

# DISCUSSION

Our findings provide a large-scale and standardized within-session view of EEG motor imagery decoding performance and refine how such performance should be interpreted. Three main conclusions emerge from the present work. First, the global decoding landscape was not flat: covariance tangent-space projection and Common Spatial Patterns (CSP) consistently defined the leading methodological families across datasets, in line with prior work emphasizing the strength of spatially informed approaches for EEG decoding [56,72–74]. Second, these aggregate rankings concealed substantial subject-level heterogeneity, as the identity of the best-performing configuration varied markedly across individuals. Third, this heterogeneity was not only descriptive but practically exploitable: compact portfolios of pipelines derived from the exhaustive benchmark retained a substantial fraction of the subject-specific oracle while reducing the effective search space.

At the benchmark level, the present results reinforce the central role of spatially informed EEG representations in motor imagery decoding. Across datasets, covariance tangent-space projection and CSP were the only families that consistently occupied the leading positions, whereas connectivity-based tangent-space representations and nonlinear descriptors remained globally secondary. This general pattern is broadly consistent with prior literature highlighting the robustness of covariance-based and Riemannian approaches in BCI decoding [32,33,56,73–75] and with MOABB-related reproducibility work suggesting that spatially structured representations remain highly competitive across BCI benchmarks [76–79]. However, the present results also show that this hierarchy is not fixed. Cho2017 favored covariance tangent-space representations, Zhou2016 favored CSP, and PhysionetMI placed the two families in near-equilibrium. Thus, even among the strongest approaches, performance depended on the interaction between methodological family, dataset characteristics, and frequency band, rather than on a single universally dominant decoding principle.

The most important conceptual result of the study is the extent of subject-level heterogeneity. Although benchmark-level averages suggested a relatively stable leading group, the subject-level analyses showed that the best-performing feature family, complete pipeline, classifier, scaler, and frequency band could all vary substantially across individuals. This finding is consistent with broader evidence that EEG signals are strongly shaped by stable inter-individual factors as well as by fluctuations in state and recording conditions [13,34,63,80–82]. Functional connectivity itself has been reported to be strongly subject-specific, and cross-dataset or cross-subject degradation has repeatedly been identified as a major challenge for EEG decoding [13,34,63,82,83]. In that context, our benchmark suggests that high-performing EEG decoding should not be interpreted as the discovery of a single universally optimal pipeline, but rather as the identification of families of competitive solutions whose relative advantages depend on the subject and the dataset.

This interpretation also provides a useful perspective on the longstanding issue commonly referred to as “BCI illiteracy” [10,19,67]. Our results do not only view poor performance necessarily reflecting an intrinsic inability of some users to generate informative motor imagery signals. They also suggest that in at least some cases, low performance may arise from a mismatch between the subject’s neurophysiological profile and the decoding pipeline being applied. The fact that nonlinear or connectivity-related approaches outperformed the globally dominant families for some individuals supports the idea that performance limitations cannot always be attributed to the user alone. This interpretation is compatible with recent arguments that so-called BCI illiteracy may partly reflect the limitations of the algorithmic interface rather than a stable inability of the participant [20].

The portfolio analysis extends this point in a practically relevant direction. Exhaustive benchmarking provides a detailed and reproducible empirical landscape, but it is computationally demanding and difficult to translate directly into deployment settings if very large numbers of pipelines would need to be evaluated for each new user. By using the benchmark itself as the basis for portfolio construction, we were able to test whether a compact subset of pipelines could preserve most of the subject-specific oracle. The answer was strongly dataset-dependent. In Cho2017, compact portfolios yielded modest but consistent gains over the best single global pipeline. In PhysionetMI, the gains were substantially larger and continued to increase with portfolio size, consistent with the broader and flatter optimization landscape observed in this dataset. In Zhou2016, by contrast, the room for improvement was negligible, which is coherent with its high baseline performance and limited heterogeneity. These results suggest that the practical value of portfolio construction is directly linked to the amount of subject-level heterogeneity present in the decoding landscape.

A second important insight from the portfolio analysis is that the most useful diversity did not primarily arise from broad mixing across unrelated feature families. Instead, the selected portfolios were usually dominated by variants of the globally strongest family already identified by the benchmark: cov-tgsp on Cho2017 and PhysionetMI, and CSP on Zhou2016. In other words, the benefit of portfolio construction came less from combining fundamentally different categories of methods than from exploiting complementary classifier or preprocessing variants within a strong methodological family. This is an important nuance for practical BCI design. It suggests that reducing the search space may not require highly eclectic ensembles, but may instead be achievable through carefully selected subsets within the families that already perform best on average. The weak performance of the family-constrained strategy is fully consistent with this interpretation.

The comparison across portfolio-construction strategies further strengthens this conclusion. Although topk_mean provided the strongest global trade-off and was therefore retained for the main analysis, regret_greedy, greedy, and beam yielded closely related results. The differences between the strongest strategies were small relative to the much larger effects associated with dataset heterogeneity and portfolio size. This indicates that the central result of the portfolio analysis is not an artifact of one specific search rule. Rather, it reflects a genuine structural property of the benchmark: when subject-level heterogeneity is sufficiently large, compact sets of pipelines can recover much of the subject-specific oracle; when heterogeneity is limited, the room for improvement over the best single global pipeline is naturally reduced.

These findings should nevertheless be interpreted in light of important limitations. First, the benchmark was conducted under a within-session MOABB protocol and therefore evaluates generalization to unseen trials from the same recording session only. It does not test cross-session robustness, long-term non-stationarity, or participant-independent transfer, all of which remain major challenges for real-world BCI deployment [9,34,63,81,82]. Second, the portfolio analysis is benchmark-driven rather than online or prospective: it shows that compact subsets can preserve much of the subject-specific oracle under repeated subject-level train-test splits, but it does not yet provide an operational mechanism for selecting the best pipeline for a truly new user in real time. Third, the present study focuses on three motor imagery datasets and on a large but still predefined space of classical pipelines, preprocessing procedures, and shallow neural classifiers. It therefore maps an extensive empirical landscape, but not the entire space of possible decoding strategies. Finally, Zhou2016 contains only four participants and must be interpreted cautiously, particularly when discussing the apparent saturation of the optimization landscape.

Several directions follow naturally from these results. A first priority is to examine whether the same benchmark-driven portfolio logic remains useful under more stringent protocols, especially cross-session and participant-independent evaluation. A second is to couple compact portfolios with lightweight subject-selection mechanisms, such as calibration-based ranking, transfer learning, or domain alignment procedures, in order to choose among the pipelines contained in the portfolio without requiring exhaustive evaluation. A third is to extend the same benchmarking and portfolio framework to additional paradigms and modalities, including P300, SSVEP, CVEP, invasive recordings, and multimodal combinations with EMG, EOG, or other biosignals [4,23]. Such extensions would help determine whether the present combination of strong dominant families, marked subject-level heterogeneity, and partial recovery through compact portfolios is specific to motor imagery EEG or reflects a broader property of neurophysiological decoding.

The present study also has implications for future methodological development. Deep learning remains promising for EEG decoding [22–24,28], but its performance and stability often depend on larger training resources and careful optimization. In this context, benchmark-driven portfolio construction may provide a useful intermediate strategy between fully exhaustive evaluation and fully adaptive subject-specific modeling. Rather than requiring immediate deployment of highly complex adaptive systems, one can first identify compact sets of candidate pipelines that cover most of the empirically relevant variability. Such an approach may offer a practical bridge between classical benchmark science and more flexible personalized BCI systems. Another next step is not only to construct compact portofolios, but to learn a subject-to-pipeline matching function grom benchmarked subjects and their features profile.

Overall, our results argue against a one-size-fits-all view of EEG motor imagery decoding. Covariance tangent-space projection and CSP remain the strongest global candidates, but their benchmark-level dominance does not translate into a single universally optimal end-to-end configuration. Instead, the data support a more nuanced interpretation in which robust EEG decoding requires both identifying leading methodological families and acknowledging substantial subject-level heterogeneity within them. From this perspective, exhaustive benchmarking is not only useful for ranking methods; it can also serve as an empirical basis for reducing the practical search space and constructing compact subject-aware sets of candidate pipelines. This shift from exhaustive evaluation to benchmark-driven portfolio selection may provide a useful framework for connecting methodological rigor with deployable personalization in future BCI research.

Finally, as personalized EEG decoding moves closer to clinical and real-world deployment, ethical considerations must remain central [84–86]. More adaptive and subject-aware systems will likely require richer individual data, repeated calibration, and potentially multimodal profiling, all of which raise important questions about privacy, transparency, fairness, and responsible use of neural data. Technical advances in personalization should therefore be accompanied by equally careful reflection on the ethical and legal conditions of deployment.

## CONCLUSION

This study provides a large-scale, standardized benchmark of EEG motor imagery decoding pipelines across three publicly available datasets and shows that decoding performance should not be interpreted solely through average rankings. At the global level, covariance tangent-space projection and Common Spatial Patterns emerged as the most competitive methodological families, but their relative performance remained dataset-dependent. At the subject level, however, this apparent hierarchy fragmented substantially: the best-performing configuration varied across individuals in terms of feature family, preprocessing strategy, classifier, and frequency band. These results indicate that no single end-to-end pipeline can be considered universally optimal, even within a controlled within-session setting.

By extending the benchmark into a portfolio analysis, we further showed that this heterogeneity is not only descriptive but practically exploitable. Compact portfolios of pipelines, selected directly from the exhaustive benchmark, retained a substantial fraction of the subject-specific oracle while reducing the effective search space. The value of this reduction was strongest in the most heterogeneous dataset and negligible in the most saturated one, indicating that the practical benefit of pipeline reduction depends directly on the structure of the underlying decoding landscape. Importantly, the useful diversity captured by these portfolios arose mainly within the strongest methodological family rather than from broad mixing across unrelated families.

Taken together, these findings argue against a one-size-fits-all view of EEG motor imagery decoding and support a more nuanced framework in which robust performance depends on both identifying globally competitive methodological families and acknowledging substantial subject-level heterogeneity within them. In this perspective, exhaustive benchmarking is not only a tool for method comparison; it can also serve as an empirical basis for constructing compact, subject-aware sets of candidate pipelines. This benchmark-driven transition from exhaustive evaluation to practical pipeline reduction may provide a useful framework for future BCI research seeking to connect methodological rigor, reproducibility, and deployable personalization.

Future work should test whether this logic remains valid under cross-session and participant-independent protocols, extend the same approach to additional BCI paradigms and multimodal recordings, and integrate lightweight subject-selection mechanisms capable of identifying the most appropriate pipeline from within a compact portfolio. More broadly, the present results support a shift in emphasis for EEG-based BCIs: from the search for a single universal decoder toward strategies that explicitly account for the structured variability that makes each user, dataset, and recording context different.

## DATA AVAILABILITY STATEMENT

The open datasets analyzed during this study are publicly available via the MOABB repository, including:

- the PhysioNet Motor Imagery dataset https://moabb.neurotechx.com/docs/generated/moabb.datasets.PhysionetMI.html;
- the Cho2017 dataset

https://moabb.neurotechx.com/docs/generated/moabb.datasets.Cho2017.html;
- the Zhou2016 dataset https://moabb.neurotechx.com/docs/generated/moabb.datasets.Zhou2016.html.

## FUNDING

This research received no specific grant from any funding agency in the public, commercial, or not-for-profit sectors. The salaries of the researchers were covered through their respective institutional and/or corporate employers.

## CONFLICT OF INTEREST

Olivier Oullier and Paul Barbaste are co-founders of *Inclusive Brains*, which maintains an active research partnership with *IBM*. Xavier Vasques is Vice President and Chief Technology Officer of *IBM Technology France*. The Inclusive Brains x IBM research partnership includes research on artificial intelligence, neurotechnologies and quantum informatics. However, it did not influence the design, data collection, analysis or interpretation of the present study. The research reported here was carried out for academic purposes.

Paul Barbaste is Senior Consultant at *Wavestone*. Olivier Oullier is Professor at the *Mohamed bin Zayed University of Artificial Intelligence* (MBZUAI), CEO of *Inclusive Brains* and Chairman of *Biotech Dental Group*'s AI Institute. He is also a shareholder and former President of *EMOTIV Inc*. Xavier Vasques is Head of Laboratory and Mathematician at the the *Institut du Neurone*.

No additional financial or non-financial competing interests are declared by the authors.

## Acknowledgments

We thank Dr. Laura Cif for helpful discussions and insightful suggestions.